# Extra-Low-Frequency Magnetic Fields alter Cancer Cells through Metabolic Restriction

Ying Li[1,2] and Paul Héroux[1,2]

[1] InVitroPlus Laboratory, Department of Surgery, Royal Victoria Hospital, Montreal, Canada.
[2] Department of Epidemiology, Biostatistics and Occupational Health, McGill University, Montreal, Canada.

Contact: Dr. Paul Héroux, Faculty of Medicine, McGill University, Montreal, Canada, H3A 1A3.  1-514-398-6988     paul.heroux@mcgill.ca.

**Abstract**

BACKGROUND: Biological effects of extra-low-frequency (ELF) magnetic fields (MF) have lacked a credible mechanism of interaction between MFs and living material.
OBJECTIVES: Examine the effect of  ELF-MFs on cancer cells.
METHODS: Five cancer cell lines were exposed to ELF-MFs within the range of 0.025 to 5 µT, and the cells were examined for karyotype changes after 6 days.
RESULTS: All cancer cells lines lost chromosomes from MF exposure, with a mostly flat dose-response. Constant MF exposures for three weeks allow a rising return to the baseline, unperturbed karyotypes. From this point, small MF increases or decreases are again capable of inducing karyotype contractions. Our data suggests that the karyotype contractions are caused by MF interference with mitochondria's ATP synthase (ATPS), compensated by the action of AMP-activated protein kinase (AMPK). The effects of MFs are similar to those of the ATPS inhibitor oligomycin. They are amplified by metformin, an AMPK stimulator, and attenuated by resistin, an AMPK inhibitor. Over environmental MFs, karyotype contractions of various cancer cell lines show exceptionally wide and flat dose-responses, except for those of erythro-leukemia cells, which display a progressive rise from 0.025 to 0.4 µT.
CONCLUSIONS:  The biological effects of MFs are connected to an alteration in the structure of water that impedes the flux of protons in ATPS channels. These results may be environmentally important, in view of the central roles played in human physiology by ATPS and AMPK, particularly in their links to diabetes, cancer and longevity.

| | |
|---|---|
| AMPK | Adenosine MonoPhosphate Activated Protein Kinase |
| ATPS | Adenosine TriPhosphate Synthase |
| ELF | Extra-Low-Frequency |
| KC | Karyotype Contraction |
| MF | Magnetic Field |

**Introduction**

Since the 1979 Wertheimer and Leeper [1] article linking wire codes to childhood cancer, the relation between cancer and power-frequency magnetic fields (MFs) has been under investigation [2]. Population, *in vivo* and *in vitro* studies have failed to provide a clear link. The exception is childhood leukemia [3], leading the International Agency for Research on Cancer to attach the class 2B carcinogen designation to MFs in June 2001 [4].

It has been argued that environmental 60-Hz MFs, as non-ionizing radiation, and incapable of raising tissue temperatures, could not have significant impacts on cells. But effects on breast cancer cells MCF-7 were confirmed by a number of laboratories near 1.2 µT [5]. Many have also reported a diversity of effects above 2.5 µT, higher than common environmental exposures. These include lengthened mitotic cycle and depressed respiration [6] increased soft agar colony formation [7], inhibition of differentiation with increased cell proliferation [8], as well as DNA breaks with apoptosis and necrosis [9].

In the early days of ELF MF research, Semikhina at al [10; 11] documented by electrical dissipation factor ($\omega RC$, also known in electrical engineering as *tg δ*) and optical measurements (the dimerization of dilute rhodamine 6G solutions) that alternating MFs in the range 25 nT- 879 µT disrupt the arrangement of water molecules, particularly under *high concentrations of hydrogen bonds and protons*. The effects were absent above 40-50°C, as water structure changes. The maximum effect was detected at 156.2-Hz and 15.45 µT for 7°C pure water. Narrow resonances were observed, easily broadened by the presence of even small levels of impurities. The MF effects on water progressed over 5 hours, and dissipated over 2 hours after the field was turned off.
Interestingly, when alternating MFs were kept below 25 nT, an influence of static MFs on water could be detected. Removing the static MF acted on water variables (dissipation factor and optical measurements) in a direction opposite to the application of ELF MFs larger than 25 nT. Thus, it seemed that elimination of both ELF and static MFs allowed water to 'optimize' its molecular arrangement.

These observations created ground to attempt an interpretation of ELF MF health effects based on water structure alterations brought about by the MF itself, as opposed to magnetically-induced currents. We investigated this possibility by setting up baseline cancer cell lines maintained under power-frequency MFs lower than 4 nT, and also under anoxia.

As 82 % of oxygen readings in solid tumours are less than 0.33 % [12], and stem cells are hosted in niches that are very low in oxygen [13], anoxia is a better simulation of the tumour environment than routinely used 21 % oxygen. Our cells are also hyperploid, displaying a range of chromosomes numbers larger than 46, as a result of the enhanced metabolism typical of cancer cells. The absence of oxygen reduces chromosome numbers to some extent, but not back to normal, and also narrows their range [14].

*Metabolic restrictors,* chemicals that impair oxygen metabolism, ATP synthesis or ATP use, can bring back chromosome numbers in cancer cell lines even closer to their original 46 than anoxia, an effect labelled *karyotype contraction* (KC). KC is a rapid and reversible loss of chromosomes resulting from metabolic restriction [14].

A critical enzyme in ATP production is ATPS. The structure of ATPS is documented in detail [15] as a rotating motor-generator structure activated by the trickle of high-density protons from the inter-membrane space into the matrix of mitochondria. Proton diffusion along the 15 nm thick inter-membrane space does not limit their transit time of 1 to 2 µs [16]. Protons enter the Fo of ATPS along an *entry half-channel* made of four hydrophilic α- helices, to reach a rotating helix. After rotation, protons flow out through a similar *exit half-channel*. The rotation is used by the $F_1$ segment of ATPS to produce ATP [17].

These hydrophilic channels [18] provide a high density of hydrogen bonds, while the mitochondrial inter-membrane space feeds ATPS a high-density of protons. The high-density protons (pH 1, [16]) are driven through the half-channels by a 180 kV/cm electric field [19] across the inner membrane [20].

In this study, we assess the ability of MFs at common environmental levels to induce KCs.

## Results

Because of the controlled MFs and of anoxia, our reference K562 cultures are karyotypically and otherwise exceptionally stable. 75 % of the cells have just two chromosome numbers, 62 and 61, compared to a wider range under 21 % oxygen [14]. The stability of chromosome numbers in baseline anoxic K562 has been periodically confirmed in our lab over 5 years. These cultures provide an extremely precise reference point, as shown in the narrow baselines of Figs. 1, 2 TOP and 4. This is of great advantage in obtaining statistical significance in our data. Fig. 1 shows little overlap between baseline and exposed data, yielding small numbers in Student's t-tests. In Fig. 2 TOP, the p-value between baseline and 0.025 µT is 0.00012. In Fig. 3, even when using 21 % oxygen, the large number of karyotypes performed and the strong shifts in average chromosome numbers produced by MFs result in extremely small p-values (0.000006 for MCF7).

### *Induced Currents*

Whether biological effects of power-frequency MFs relate to the MF itself, or to the currents induced in tissues by the fields, has been a perennial question. Many think that effects occur through potentials produced by magnetically induced currents blocked by the thin membranes, within or bordering living cells. Such currents and membrane potentials are familiar to conventional electrophysiology.

In the results of Fig. 1, one aliquot of an anoxic K562 cell culture is placed in a vertical, and the second in a horizontal MF exposure system. At the same MF, the horizontal coil induces currents 6 times larger because the exposed culture dish area is 34 x 34 mm for the horizontal coil, compared to 5.8 x 34 mm for the vertical coil. As KCs after 6-days at 1 µT come out similarly for both orientations (Fig. 1), we conclude that the effect on chromosome numbers are dependent on the MF itself. We assume direct MF, rather than induced current action on the basis that variations of current density by a factor of 6 do not affect KC. But this would also occur if induced currents had a flat dose-response, already saturated at the lower current. Further, direct MF action on KC does not preclude that other effects of MFs may depend on induced currents.

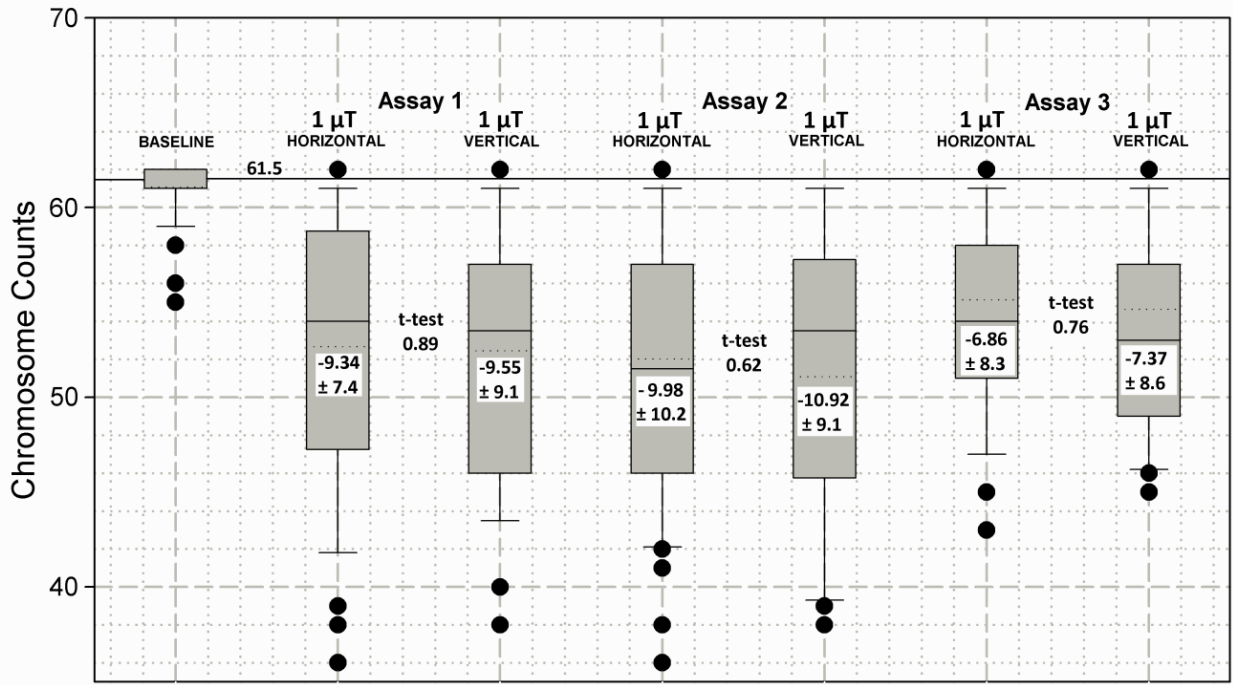

**Fig. 1.** Baseline anoxic K562 cells at less than 4 nT (60-Hz) with an average of 61.5 chromosomes (horizontal line), and a very narrow distribution (at left) are simultaneously transferred for 6 days to 1 µT MFs applied either horizontally or vertically. Three independent 6-day assays show the resulting chromosome numbers. Box plots show median (solid), average (dotted), 25 and 75 % (box), 10 and 90 % limits (whiskers), and outside values (dots). 56 (Assay 1), 50 (Assay 2) and 51 (Assay 3) metaphases were karyotyped in each orientation. Inside the box plots are average chromosome losses. The Student's t-test results quantify the probability that the horizontal and vertical results are identical.

### *Dose-Response*

Fig. 2 TOP shows the chromosome number losses experienced by previously shielded anoxic K562 cells after 6-days in various MFs. Under any exposure, the narrow baseline expands, and there are substantial KCs.

Three features are of importance. First, a no-effect-level lower than 25 nT. Second, a progression of KCs up to 0.4 µT. Third, the relatively flat dose-response between 0.1 and 1.5 µT.

The graph spans time-averaged MFs representing domestic (0- 0.2 µT), commercial (0.07- 0.5 µT) and occupational (0.1-1µT) environments [21].

### *Across Cell Lines*

Beyond K562, we investigated four more hyperploid cancer cell lines to determine the generality of KC by MFs. Over two orders of MF magnitude, erythro-leukemia (HEL 92.1.7), breast (MCF7) and lung (NCI-H460) cancer cells lose between 8 and 13 chromosomes (Fig. 3). HEL, our second erythro-leukemia cell line, shows fewer losses at lower fields, similar to K562. Three of the four results reported in Fig. 3 were obtained under standard (21 % oxygen) culture conditions.

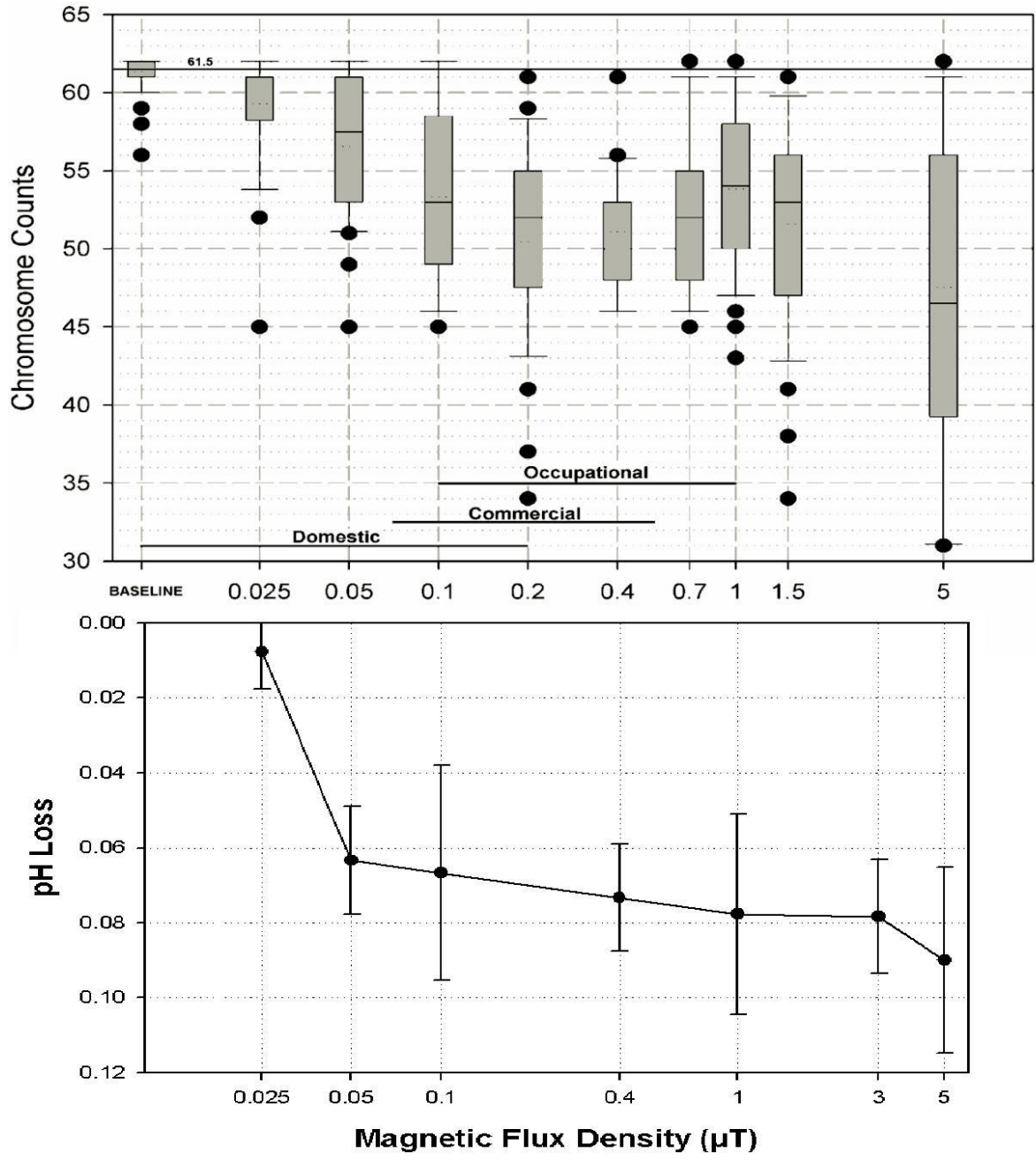

**Fig. 2. TOP:** K562 chromosome numbers as a function of 60-Hz Magnetic Flux Density. 6-day assays with, in sequence, 65, 28, 50, 77, 46, 33, 65, 102, 56 and 50 metaphases. 2 to 6 experiments at each MF. Approximate ranges for domestic, commercial and occupational exposures are shown.

**BOTTOM:** pH differences between two cell medium aliquots, one exposed for 20 hours to <4 nT at 60-Hz, and the second to the MF density in the Figure. Medium is RPMI-1640 with 10 % FBS. Isotherm measurements using Auto Read were made with the same probe, alternating between the two aliquots. 3 measurements for each aliquot, and 3 repeats at each MF density.

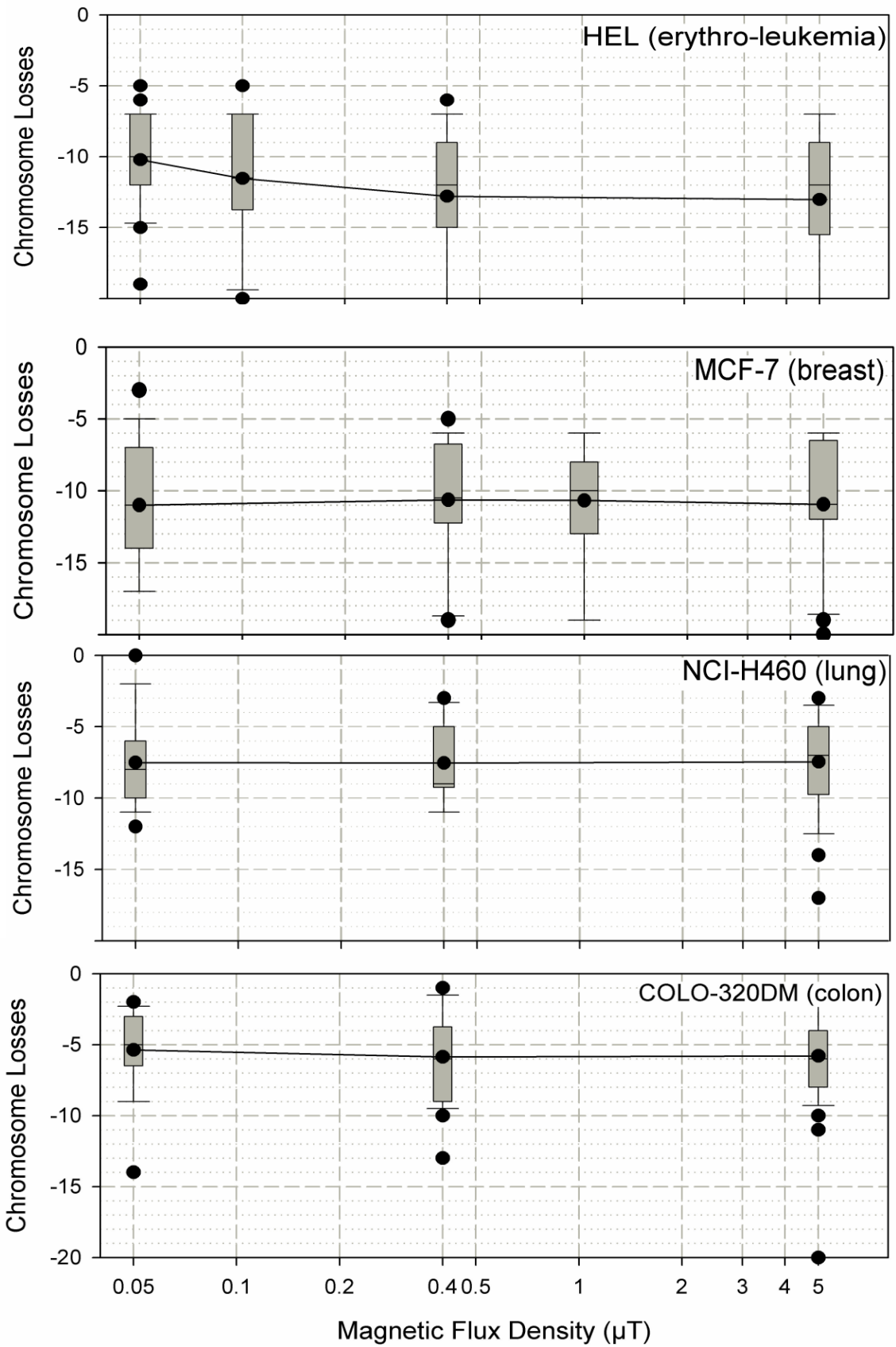

**Fig. 3 AT LEFT.** Average chromosome losses in erythro-leukemia, breast, lung and colon cancer cells as a function of 60-Hz Magnetic Flux Density. The chromonome number baseline ("0") averages for < 4 nT cells at 60-Hz, 80 % range and metaphase number are: HEL: 66, 62-67, 32; MCF7: 74, 61-75, 30; NCI-H460: 57, 53-65, 30 and COLO 320DM: 54, 49-61, 30. 6-day assays with, in sequence, 32, 22, 29, 32; 19, 22, 19, 21; 29, 22, 24; 22, 34 and 46 metaphases. 2 to 5 experiments at each MF. HEL, NCI-H460 and COL 320DM assays used 21 % oxygen, rather than anoxic conditions, as some anoxic karyotype modes are too close to 46 to allow easy statistical separation from MF-exposed samples.

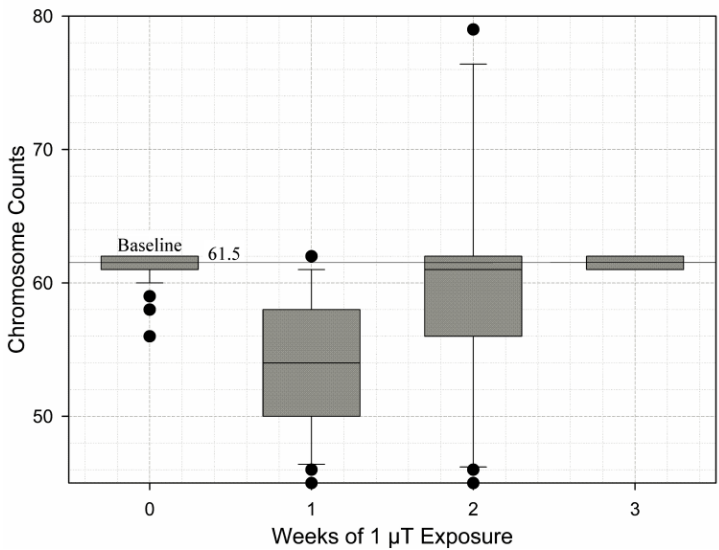

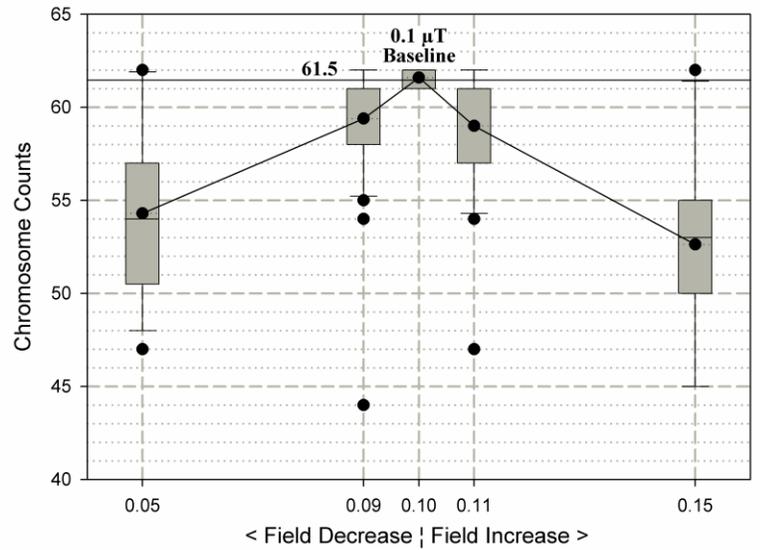

**Fig. 4. LEFT:** K562 chromosome numbers return to baseline after 3 weeks of continuous 1 µT MF exposure. 65, 102, 50 and 37 metaphases. 2 experiments at each MF.

**TOP RIGHT:** K562 Chromosome numbers obtained after 6 days by altering baseline MFs of 0.1 µT. 20, 31, 37 (baseline), 31, 35 metaphases. 3 to 6 experiments at each MF.

**BOTTOM RIGHT:** For 1 µT, 28, 28, 37 (baseline), 28, and 28 metaphases. 3 experiments at each MF. Although the symmetry of the chromosome numbers is strong, there is more cell decay with increased than with reduced fields.

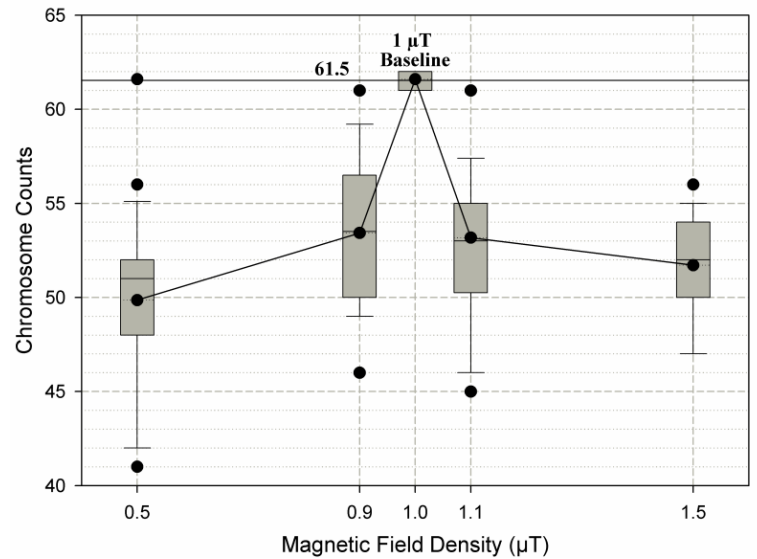

Classical toxicology and epidemiology, where smoothly climbing dose-responses are justified by binding chemistry and the central tendency theorem, do not expect the flat dose-responses observed in Fig. 2 TOP and Fig. 3. The effects found for different cell types are strikingly similar, with similar low-field deviations in the two erythro-leukemia lines, suggesting common, basic mechanisms.

### *Differential Action*

K562 cells with magnetic KCs such as in Fig. 2 TOP progressively recover their original chromosome numbers after 3 weeks, even as the MF is maintained at a constant level (Fig. 4 LEFT). Surprisingly, in cells recovering over 3 weeks from a MF disturbance, the deviation of chromosome numbers is even less than what is observed in the long-term baseline culture, as shown in the last measurement of Fig. 4 LEFT and in the central measurements of Figs. 4 RIGHT TOP and BOTTOM. Chromosome numbers restore earlier than chromosome number dispersions.

After 3 weeks, if the MF is altered by a small percentage of the original value, either positively or negatively, KCs are again observed, as shown in Fig. 4 RIGHT. Starting from low (TOP, 0.1 µT) or high (BOTTOM, 1 µT) baselines, symmetrical KCs are observed. This bilateral sensitivity to changes is unforeseen by conventional toxicological principles. KC is also observed when fields are reduced from 50 to 4 nT (not shown).

The KCs will be interpreted below as caused by magnetically-induced perturbations in intra-cellular ATP levels. These results cast doubt on the stability of cancer cell models housed in incubators with MFs that are highly variable over space and time [22].

### *Over Frequency*

We measured in anoxic K562 6-day tests at 1 µT the average KCs over frequency as follows: -3.6 ±0.79 at 50-Hz, -9.36 ±1.06 at 60-Hz, -12.71 ±1.82 at 120-Hz and -9.8 ±1.31 at 155-Hz. A polynomial fit predicts maximum KC effect on ATPS at 113 Hz for 1 µT.

### *Static Magnetic Field Removed*

The influence of the static MF was investigated by observing K562 cells transferred from a steel shield that eliminated ELF MFs (to less than 5 nT), but had a static field of 74 µT, to a second shield ('NIM') that attenuated both the ELF MF (less than 5 nT) and the static field to 3 µT. Karyotyping revealed a very slow drift downward, but a strong effect on proliferation rate was observed. After 4 days, cell numbers in the NIM shield were increased by a factor of 2.05 ± 0.13 (S.D.) over cells kept in the steel shield, indicating enhanced metabolism. The effect is persistent over time.

### *Magnetic Field and Oligomycin*

Previous experiments [14] on the 5 cancer cell lines used in this article show a link between metabolic restriction and KC. Anoxia alone induces partial KCs of 6-8 chromosomes. Deeper contractions, almost to normalization of the karyotypes to 46, are produced by IC50 doses (allowing 50 % of the normal cell division rate) of the metabolic restrictors oligomycin and imatinib. Similar KCs are produced by physiological levels of melatonin and vitamin C together. We believed that comparison of metabolically restricted cultures with MF-exposed cultures could provide clues on action mechanisms, as the different metabolic restrictors mentioned above have different sites of action. Fig. 5 TOP, displays the similarity in cell size distribution after 6-days between two of seven anoxic K562 assays, one exposed to a very effective MF, 0.4 µT at 60-Hz, and the second to oligomycin at IC50 (2.5 ng/ml). The two distributions stand apart, with smaller cell diameters and higher ratios of cells-to-objects below 11 µm, the decay particles and apobodies. This suggests that MFs and oligomycin share a common mode of action. Despite the closeness between oligomycin and MF assays in Fig. 5 TOP, oligomycin is faster-acting than 0.4 µT: changes in cell size, revealing of KC, are visible at 1 day, earlier than for the MF. But more efficant MFs, such as 5 µT at 60-Hz or 1 µT at 120-Hz, show effects earlier (not shown).

### *Magnetic Field and AMPK*

The similarity between 0.4 µT and oligomycin suggests that the MF may be an inhibitor of ATPS, as oligomycin is a highly specific inhibitor of ATPS. If this were the case, inhibition of mitochondrial ATPS by MFs would activate AMPK, because healthy cells must maintain a high level of phosphorylation capacity (ATP:ADP ≈ 10) to function well [23]. AMPK is a sensitive ATP regulator that switches *on* catabolic pathways and *off* many ATP-consuming processes, both acutely and chronically, through gene expression.

The MF>ATPS>AMPK pathway was investigated using metformin and resistin. Metformin is a diabetes drug that activates AMPK, leading to reduced glucose production in the liver, and reduced insulin resistance in muscle. It is an attractive anti-aging drug that usually causes weight and appetite loss.

Resistin, a product of the RSTN gene, is a 9.9 kDa protein containing 93 amino acid residues which, at 20 ng/ml or more, inhibits AMPK. It interferes with phosphorylation of Akt (serine/threonin protein kinase), active in multiple cellular processes such as glucose metabolism, cell proliferation, apoptosis, transcription and cell migration.

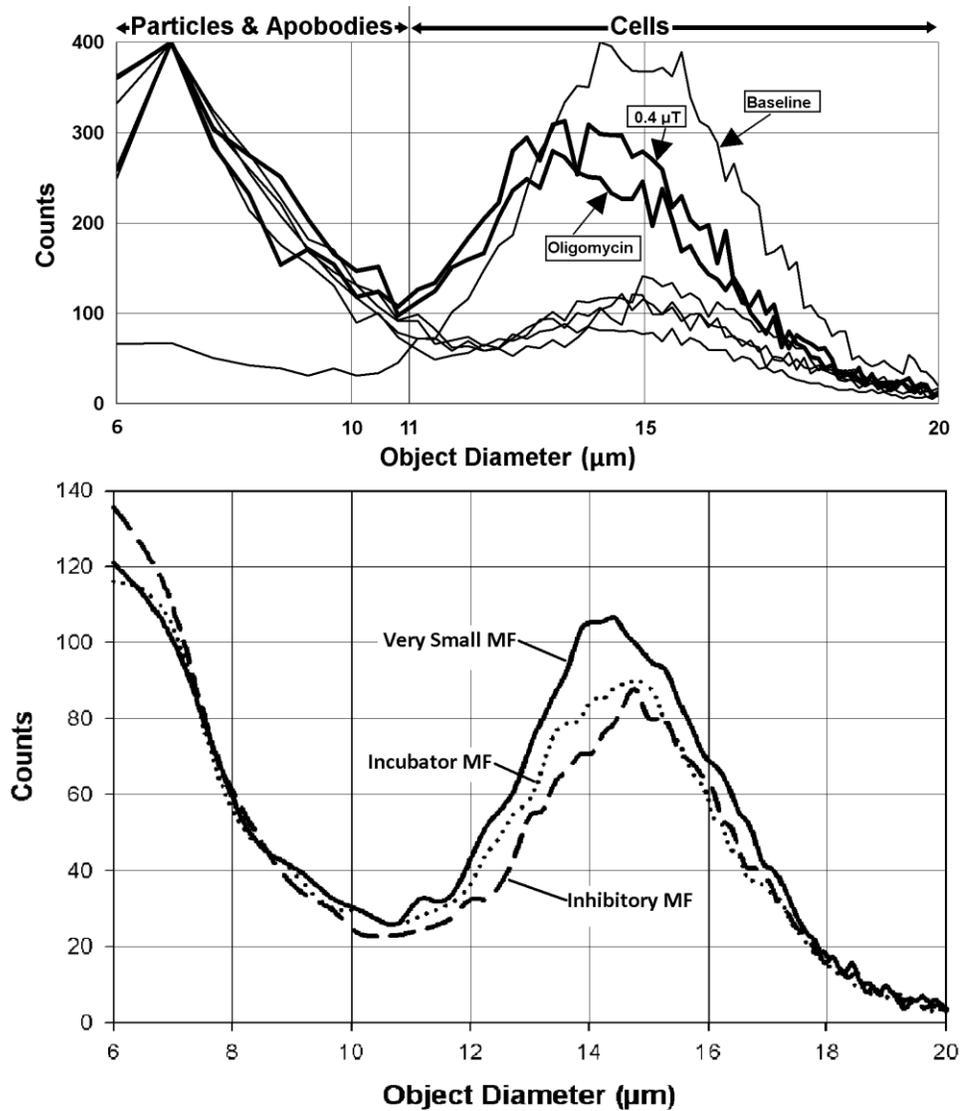

**Fig. 5. TOP: Object Diameter histograms for 6-day anoxic exposures of K562 cultures to 0.4 µT MF at 60-Hz and oligomycin at IC50 (2.5 ng/ml). The lower 4 IC50 curves are, bottom to top at 15 µm: imatinib (0.04 µg/ml), resistin (40 ng/ml), metformin (0.01 mg/ml) and melatonin-vitamin C (0.3 µg/ml, 26 µg/ml ). All cultures are adjusted to a common small particle count maximum.**

**BOTTOM: Object Diameter histograms for 7-hour 21 % oxygen exposures of three K562 cultures under typical Incubator MF. Aliquots of RPMI-1640 with 10 % FBS medium exposed for 15 hours to Very Small MF (<4 nT at 60-Hz, 3 µT static), Incubator MF (2 to 2.7 µT at 60-Hz) or Inhibitory MF (0.62 µT at 120 Hz) were seeded with cells at time 0, and measured with a Millipose Scepter at 7 hours. Average of 3 repeats for each condition, The p-value between average levels (12-16 µm) for Very Small MF and Inhibitory MF is 0.001 (n=4).**

Metformin (0.01 mg/l) and resistin (40 ng/l) alone for 3 days induce average KCs of 9 and 10 respectively, in K562. When, in the 6-day trials routinely used for MF tests, 1 µT is added to metformin, even *larger* KCs are observed (9 becomes 11 ± 0.34). When 1 µT is added to resistin, the KC of resistin *reduces* from 10 to 4 ± 0.46, also less than the KC of 1 µT alone, at 7.5. The conclusion is that MFs enhance the action of metformin, but neutralize the effect of resistin, again suggesting a connection between MFs and ATPS.

### *Experiments on Medium Alone*

Cells grown for 7 hours under identical *Incubator MF* (2 to 2.7 µT at 60-Hz) conditions fared differently according to whether the culture medium added at 0 hours originated from closed flasks exposed for 15 hours to: *Very Small MF* (<4 nT at 60-Hz, 3 µT static), *Incubator MF* (2 to 2.7 µT at 60-Hz) or *Inhibitory MF* (0.62 µT at 120 Hz).

After the sealed flasks with media (only) are exposed to their respective MFs, cell culture aliquots are introduced into each flask, and incubated for 7 hours under *Incubator MF* conditions. Measurements of cells numbers of each size are acquired at 0 hours, as well as at 7 hours for each flask. There is an increase (Fig. 5 BOTTOM) in the number of living cells observed under the *Very Small MF* condition, compared to the *Inhibitory MF* condition, with the *Incubator MF* condition rating in between.

When stressed cells from a culture with depleted medium (lower pH) were used, the *Inhibitory MF* had the effect of increasing the level of decay products (object diameters less than 11 µm) in the culture (not shown).

The lasting effect of MFs on aqueous fluids is also observable from pH measurements in cell culture media, which turn slightly more acidic under short MF exposures. After 20 hours, there is a difference of -0.09 pH units with a 95 % confidence interval of 0.045 between unexposed vs 5 µT 60-Hz exposed media (Fig. 2 BOTTOM) for the widely used RPMI-1640 with 10 % serum. The pH shift was confirmed for a variety of cell culture media.

### *NCI-H460 Proliferation*

Beyond strong effects on cancer cells karyotypes, MFs also impact proliferation rate, adhesion and cell shape, which cannot be reported in detail here. Some prominent effects are strongly dependant on MF intensity. For example, the cell counts of lung cancer cells (NCI-H460) after 4 days in our synthetic medium at 50 nT, 400 nT and 5 µT are 8, 9.2 and 14.8 times larger than those of unexposed cells. Unexposed NCI-H460 do not attach in our synthetic medium, but do so under any MF exposure.

## Discussion

### *Possible Biological Site of Action of Magnetic Fields*

The involvement of water structure disruptions is anticipated by recent views on EMF bioeffects [24; 25]. If the effect on water described by Semikhina and Kiselev is involved, it would be most prominent in a location where "*high concentrations of hydrogen bonds and protons*" are found. The only known location in the biota where these two conditions are met are the entry and exit water channels of ATPS.

According to this mechanism, MFs would impede proton flow through the hydrophilic channels, and MF removal would improve proton flow, directly impacting ATPS efficiency.

In this hypothesis, the dose-responses of Figs. 2 TOP and Fig. 3 are determined by rising proton impedance (decreased soliton tunneling) through ATPS half-channels. Tunnelling of protons similar to the one we hypothesize for ATPS has been observed as double wells in neutron Compton scattering studies performed on nanotubes [26].

The involvement of protons relieves the area of EMF bioeffects of the "kT problem", because EMF would act not on molecules, but on particles (protons and electrons), which, as fermions, do not follow Maxwell-Boltzmann statistics, but Fermi-Dirac statistics, and are governed by quantum electrodynamics.

Numerous elements documented by Russian physicists in their studies of MFs on water [10; 11] are compatible with our own biological observations.

It is particularly notable that the KC threshold of 25 nT in Fig. 2 TOP falls in line with the water effect threshold detected by Russian physicists [10]. The extended, flat response is also compatible with their observations. Further, the increased metabolism observed when alternating and static MFs are removed, and the ability of MF-conditioned culture media to influence cellular development are all compatible with Russian data. The presence of a KC resonance wider than that observed for pure water by Russian physicist adds support. Lastly, the ratio between frequency and field intensity (f/B) for maximum biological effects is suggestive of a coupling with the gyromagnetic ratio of the proton.

In this context, similar fingerprints between the 0.4 µT and oligomycin (Fig.5 TOP), known to inhibit ATPS by binding to the δ subunit of the Fo segment of ATPS (also named *oligomycin sensitivity conferral protein)*, comes as no surprise. Another intriguing link between MFs and ATPS is provided by the fact that rhodamine 6G, used by Semikhina to detect MF effects on water, also happens to inhibit the $F_O$ segment of ATPS.

### *Karyotype Contraction, AMPK and Diabetes*

Perturbations of ATP concentrations trigger AMPK, which activates p53 and reduces both ATP consumption and DNA synthesis [27; 28]. The suppression of DNA synthesis, part of AMPK's catabolic control, leads to KCs through suppression of chromosome endo-reduplication, the mechanism probably responsible for rapid chromosome number increases in cancer cells [14].

Two unusual aspects of MF action, adaptation to a stable field over three weeks (Fig. 4 LEFT), and the unusual shorter-term sensitivity to small MF increases and decreases (Fig. 4 RIGHT) are compatible with AMPK physiology. As far as we know, this is the first example of an agent presenting this kind of symmetry, making it possible to sustain KCs indefinitely by judicious selection of MF sequences. AMPK is easily triggered by small changes in ATP levels [23], but also controls long-term dynamic adaptation in muscle [29]. The connection between metabolic restrictors, including MFs, and KC may be be explainable by AMPK physiology.

The MF>ATPS>AMPK pathway is easily detectable in cancer cells because of KC, but there is no reason to think that the ATPS of normal cells is spared under MF exposure. A major regulator of metabolism [30], AMPK modulates insulin secretion by pancreatic beta-cells [31], and is investigated for the treatment of diabetes [32]. AMPK is tied with body weight [33] as well as with immune cell behaviour [34].

### *Karyotype Contractions and Cancer*

Cancer cells depend on glycolysis and significantly upregulate it when respiration is inhibited. The Warburg effect manifests as increased glycolysis and reduced mitochondrial respiration [35; 36]. These capabilities of cancer cells allow growth under metabolic restriction by concentration of their resources on bio-synthesis through the elimination of detoxification mechanisms associated with oxygen exposure, such as glutathione-S-transferase and CYP3A4 expression [37]. The smaller karyotypes maintained under metabolic restriction contribute to tumour core expansion, as fewer chromosomes can be more rapidly duplicated. The survival of tumours could thus be enhanced by certain levels of chronic metabolic restrictions from hypoxia, oligomycin or MFs. It has been repeatedly confirmed that cancer cells become more malignant under metabolic restriction [13; 38; 39] *in vitro* [40] and in the clinic [41; 42], to the point where it has become a central issue in tumour physiology and treatment [38]. From our data, it is logical to conclude that KC observed under metabolic restriction is a possible indicator of meta-genetic promotion in cancer cells [14].

### *Magnetic Fields and Cancer Epidemiology*

For many cancer cell types, the dose-response of KC vs MFs is remarkably flat (Fig. 3). The deviation from flatness in erythro-leukemia cells (Fig. 2 TOP and Fig. 3, HEL) is due, we suspect, to extra-mitochondrial ATP secretion in the cell membrane [43], where pH is at a physiological 7.3 rather than 1, a probable feature of this cell type [44].

If KC is indeed a marker of increased malignancy, there is a possibility of carcinogenicity from MF exposures. In such a case, the phenomenon would not be easy to document through epidemiology. First, the threshold for the effect (25 nT) is very low, which means that *all* the population is "exposed". Second, the dose-response is unusually flat (Fig. 3), such that useful low and high exposure groups with otherwise similar characteristics would be difficult to assemble. Third, the differential action of MFs may confuse conventional exposure analysis. Occupational studies are often at the forefront of epidemiological discovery because of their higher and better documented exposures. According to Fig. 2 TOP, occupational populations of low (0.1 µT) and high exposures (1 µT) have between them a KC difference of "1 chromosome". Domestic MF epidemiology on leukemia may have been successful [3; 45] because it benefited from a KC of "10 chromosomes" between 0 and 0.4 µT (Fig. 2 TOP).

The increased proliferation rates reported for lung cancer cultures may also be important. Lung cancer was pointed in at least four studies related to EMFs [46-48].

## Methods

### *Cells and Culture Conditions*

The cell lines, K562 and HEL 92.1.7 (erythro-leukemias), MCF7 (breast cancer), NCI-H460 (lung cancer) and COLO320DM (colon cancer) were obtained from ATCC. Cells are maintained under 5 % carbon dioxide and 90 % humidity, and grown in synthetic culture medium, because changes in serum can alter chromosome counts. The medium is RPMI-1640 with l-glutamine (Sigma 61-030-RM), sodium selenite 20 nM (Sigma S-5261), bovine insulin 1 mg/l (Sigma I5500), iron saturated bovine transferrin 25 mg/l (Sigma T1408), sodium bicarbonate 2 g/l (Sigma S-6014) and bovine serum albumin 4 g/l (Sigma A3311). Vented T-25s (Sarstedt 83.1810.502) and T-12s (Falcon 353018) were used for experiments, and cells are seeded at 5000/cm², and kept in the same medium for 6 days. In longer tests (3 weeks), new medium is added weekly. Oxygen was eliminated by enclosing T-25s and T-12s in large polycarbonate containers (1.6 L ) flushed with medical grade nitrogen (95%) and $CO_2$ (5%). pH readings were conducted under isothermal conditions (water bath) for samples as well as calibration buffers, using Corning 445 meters.

### *Magnetic Fields*

Unexposed cells for experiments are kept in T-12 or T-25 culture flasks under anoxia and MFs below 4 nT. Three 6.3 mm thick layers of structural steel reduce ELF MFs from incubators and the environment. Culture vessels are centered in a rectangular structural steel pipe 5.1 x 7.6 cm, itself contained in a 7.6 x 10.2 mm pipe, both 20 cm long. These two shields are placed in a 15.2 x 24.5 x 36 cm long pipe. This reduces 60-Hz MFs by a factor of 144, providing unexposed cells with a MF environment of 3 nT, slightly below the measurement floor (5 nT at 60-Hz) of our Narda EFA-300 instrument. The incubator is a Forma 3310, with low average MF (0.4 µT). MFs are applied by rectangular coils (19 x 25.6 cm) with 20 to 50 turns of #25 AWG varnished copper wire wound on 13 mm polycarbonate, providing $\approx 8\ \Omega$. The coil is under the two inner shields and over an acrylic spacer at the bottom of the outer shield. 60-Hz fields above 0.4 µT are from sector-connected variable transformers fitted with a passive low-pass capacitive filters, with all harmonics at less than – 20 dB. Smaller 60-Hz fields and other frequencies were generated with computer-based synthesizers with a background noise at less than – 40 dB. MFs are within 10 % of nominal in the whole cell culture area.

The "NIM" shield cancelling both alternating and static MFs is an acrylic cylinder 5.7 cm in internal diameter with a 0.38 cm wall and 38 cm in length, covered by 6 layers of 0.4 mm Nickel-Iron-Molybdenum foil (ASTM A753 Type 4) wound in a spiral, together with a 1.6 mm neoprene membrane spacer.

60-Hz 5 µT exposures produce no measurable temperature rises. K562 is a good thermal sentinel, hyperthermia being detectable from larger cell sizes at +0.5 K, while +1 K seriously impairs proliferation, and +2 K over a few days is lethal.

### *Chromosome, Cell and AMPK assays*

Metaphase preparation and cytogenetic analysis were performed according to the standard trypsin-Giemsa banding technique. Karyotypes are obtained using ×100 oil immersion, a Laborlux D (Leitz) microscope, and an Infinity X (21 Mpixels) CMOS camera (Lumenera). Cell proliferation and cell size histograms of are from a Scepter Automated Cell Counter (Millipore). Metformin was obtained from Sigma (D150959), and resistin from Prospec Protein Specialists, East Brunswick, New Jersey, USA.

## Conclusion

The following evidence supports inhibition of ATPS by MFs.

### *MFs alter Metabolism*
1. MFs induce KCs in 5 cancer cell lines, as do other metabolic restrictors [14].
2. MFs interact with metformin and resistin as would an AMPK activator.
3. Elimination of alternating *and* static MFs produces a durable increase in cell proliferation.

### *MFs alter Water*
4. The KC threshold (25 nT), as well as its extent two orders of magnitude, is predicted by the work of Russian physicists on water [11]. Lack of sensitivity to MF intensity or to cell type suggest the knockout of a biological enzyme by physics.
5. MF-exposed culture medium, without cells, is a vector of MF action (proliferation and cell decay).
6. Measured changes in the pH of cell culture media from MF exposures.
7. A wide KC resonance (113 Hz at 1 µT) is compatible with the work of Russian physicists on water [12].
8. KC is maximized at specific frequency-amplitude (f/B) combinations [12].

### *MFs alter ATPS Fo*
9. ATPS Fo is the only site in the biota where conditions for maximum sensitivity to MF action [22] happen together: high concentrations of protons and hydrophilic bonds in a narrow channel.
10. Strongly acting MFs induce cell culture characteristics (Fig. 5 TOP), closely matching those of a specific ATPS Fo inhibitor, oligomycin.
11. MF activation of AMPK implies a perturbation to ATP levels, thus a change in ATPS performance.
12. Rhodamine 6G, used by Russian physicists [11] to detect MF effects, is also an inhibitor of ATPS Fo.

Environmental MFs act on the core of human metabolism. Past evaluations of MF bio-effects were at a serious disadvantage because of traditional toxicological and epidemiological assumptions, that larger exposures induce larger responses. The controls of *in vitro* scientists were already randomly exposed by the MFs of their incubators. The flatness of MFs' dose-response impaired epidemiological work, as most studies, except for domestic leukemia, used tainted controls [49]. The interaction between power-frequency MFs and living cells may have been underestimated for a long time, because of these unexpected characteristics.

Some diseases appear to have strengthened, with no clear causation, as more advanced technology, in great part based on electricity, has expanded. Chronic diseases that increased or decreased in the last century, and that are connected to ATP metabolism, should be examined for a link with MFs. But our understanding of AMPK and metabolism is incomplete [50], making a link between MFs and any specific disease, such as diabetes, uncertain. MF is a physiological agonist of metformin, suggesting that MF exposure may have played a role in the increased lifespan observed in developed countries in the last century.

### Acknowledgements

To the late Nancy Wertheimer and Edward Leeper, who saw it first. To Semikhina and Kiselev, whose work made our analysis possible. We are grateful to Janet Moir and Lorne Beckman of the Royal Victoria Hospital for laboratory support, and Michel Bourdages, Institut de Recherche d'Hydro-Québec, for equipment contributions. We thank Louis Slesin for reviewing the document. The work was supported by Royal Victoria Hospital Research Institute Fund 65891. We declare no competing financial interests.